\documentclass[12pt,preprint]{aastex}
\usepackage{epstopdf}

\def\gapp{\lower 3pt\hbox{${\buildrel > \over \sim}$}}
\def\lapp{\lower 3pt\hbox{${\buildrel < \over \sim}$}}
\begin{document}

\title{Accurate Coordinates and 2MASS Cross-IDs \\ 
for (Almost) All Gliese Catalog Stars}

\author{
John Stauffer\altaffilmark{1}, 
Angelle M. Tanner\altaffilmark{2}, 
Geoffrey Bryden\altaffilmark{3,4},
Solange Ramirez\altaffilmark{3}, \\
Bruce Berriman\altaffilmark{3},
David R. Ciardi\altaffilmark{3}, 
Stephen Kane\altaffilmark{3},
Trisha Mizusawa\altaffilmark{3},
Alan Payne\altaffilmark{5}, \\
Peter Plavchan\altaffilmark{3},
Kaspar von Braun\altaffilmark{3},
Pamela Wyatt\altaffilmark{3},
J. Davy Kirkpatrick\altaffilmark{6}
}

\altaffiltext{1}{Spitzer Science Center, 
  1200 E California Blvd, Pasadena, CA 91125 }
\altaffiltext{2}{Department of Physics and Astronomy, 
  Georgia State University, Atlanta, GA 30302}
\altaffiltext{3}{NASA Exoplanet Science Institute, 
  California Institute of Technology, 770 S Wilson Ave, Pasadena, CA 91125}
\altaffiltext{4}{Jet Propulsion Lab, 4800 Oak Grove Dr, Pasadena, CA 91109}
\altaffiltext{5}{Australian National University, Mount Stromlo Observatory, Canberra, ACT, Australia}
\altaffiltext{6}{Infrared Processing and Analysis Center,
  California Institute of Technology, 770 S Wilson Ave, Pasadena, CA 91125}

\begin{abstract}

We provide precise J2000, epoch 2000 coordinates and cross-identifications to sources in the 2MASS point source catalog for nearly all stars in the Gliese, Gliese and Jahreiss, and Woolley catalogs of nearby stars.  
The only Gliese objects where we were not successful 
are two Gliese sources that are actually QSOs,
two proposed companions to brighter stars which we believe do not exist,
four stars included in one of the catalogs but identified there as
only optical companions, one probable plate flaw, 
and two stars which simply remain un-recovered.  
For the 4251 recovered stars, 
2693 have coordinates based on Hipparcos positions, 
1549 have coordinates based on 2MASS data, 
and 9 have positions from other astrometric sources.
All positions have been calculated at epoch 2000 using 
proper motions from the literature, which are also given here.

\end{abstract}


\section{Introduction}

The Gliese-Jahreiss (1991) catalog of nearby stars (also known as the CNS3) is 
the primary source list for those working on the properties of nearby
stars.  It was intended to include all stars known or thought to be located
within about 25 pc of the Sun at the time of the last update to the catalog.
Because this is a very active research topic, the catalog is now significantly
out of date from a variety of standpoints.  Most obviously, there have been
many new stars identified over the past 20 years with estimated distances
closer than 25 pc (c.f.\ Reid et al.\ 2002; Jao et al.\ 2005; Henry et al.\ 2006).  Less
obviously, the positional and other data provided in the CNS3 does not 
include new information provided in hundreds of papers published since 1989,
requiring those who are interested in these stars often to do significant
amounts of detective work in order to place each star in context.
This is particularly true for the $>$200 stars in the CNS3 catalog whose
listed positions were inaccurate by $>$30 arcseconds (and sometimes up
to $\sim$2 arcminutes), making attempts to obtain new data for these
stars particularly challenging.

As part of the work to create the NASA/IPAC/NExScI Star \& Exoplanet Database
(NStED; http://nsted.ipac.caltech.edu),
we have attempted to update one facet of the data in the CNS3.
Specifically, we have derived accurate J2000 positions for (almost)
every Gliese star, and where possible, we have cross-identified each Gliese star
to the 2MASS catalog.   In Section 2, we provide a brief history of the
Catalog of Nearby Stars.   In Section 3, we describe our methodology
for deriving J2000 coordinates for stars in the CNS3; an appendix
gives detailed information on individual stars where significant detective
work was needed. 
Section 4 compares the new positional information with older results
and presents the 2MASS photometric data as a color-color diagram with 
details given for outliers.
  
 \section{History of the Catalog of Nearby Stars}
 
 The Gliese catalog (Gliese 1957) had its beginnings in the 1950's as a list of 
stars within 20 pc of the Sun compiled by Wilhelm Gliese of the
Astronomisches Rechen-Institut (ARI) in Heidelberg.  The initial
catalog included 915 single and double stars, for a total of
1094 components.  Star numbers in the catalog were ordered by
right ascension.  Resolved double stars were designated by
the letters A and B (and C or higher if necessary).   
The second version of the catalog was published
in 1969 (Gliese 1969).  The distance limit for this version was 
increased to 22.2 pc (parallax $\geq$ 0.045$\arcsec$).  Because of
revisions to parallax estimates that occurred over time, about two
hundred stars in the 1969 catalog had final parallax estimates that were in
fact less than 0.045$\arcsec$.  There were 1529 single and multiple
stars in the 1969 catalog, for a total of 1890 components.   In
order to maintain the original Gliese star numbers and still keep
the numbering system ordered by RA, new numbers were created by adding
a decimal place - e.g. Gl 59.1 and 59.2 have right ascensions between
that for Gl 59 and Gl 60.   The next installment to the nearby
stars catalog was the Catalogue of Stars Within Twenty-Five Parsecs
of the Sun (Woolley, Epps, Penston \& Pocock 1970), which was 
described by those authors as an extension of the Gliese catalog.
New stars added by these authors were given the numbers from
9001 to 9850, with the 9000 series stars also ordered by
right ascension.  These new stars are sometimes referenced as
Wo 9xxx.  A number of the new stars in the Gliese (1969) and Woolley et al.\ (1970) catalogs have duplicate names (e.g. Gl 78.1 == Wo 9062; Gl 83.1 == Wo 9066). Next, Gliese and Jahreiss (1979, hereafter referred to as GJ79) published a
supplement to the nearby stars catalog.  Two sets of stars were
included in the supplement.  Table 1 of that paper contained 294 new stars thought
to have parallaxes $\geq$ 0.045$\arcsec$; their numbers are
GJ 1001 through GJ 1294 B, again ordered by right ascension.  
Table 2 contained 159 stars with uncertain distance data
(labeled as ``Suspected Nearby Stars'') - they were numbered 
GJ 2001 through GJ 2159.    Finally, what was labeled a
Preliminary 3rd Version of the Gliese catalog was published in
1991 by Gliese and Jahreiss (hereafter referred to as GJ91); this version is sometimes referred
to as the CNS3.  The parallax limit for this catalog was 0.039
arcseconds, corresponding to 25.6 pc.  The catalog 
contains 3803 stars, including most of the supplementary stars
discussed above, and 1388 new stars.  These new stars were
indicated as ``NN 3001'' through ``NN 4388'' in the CNS3, with the
expectation that these numbers would be superceded in the final
version of the catalog.  That did not happen, however, and these
stars are now often referred to as GJ rather than NN.  A few stars
from previous catalogs do not appear in the CNS3 (e.g. Wo 9007),
presumably because later data indicated that they did not meet
the imposed parallax limit.  In this paper we attempt to derive accurate J2000 coordinates for all stars included in any of the above mentioned nearby star catalogs (as opposed to just the stars in GJ91).
In general, if a possible binary companion was identified in one of the nearby
stars catalogs but was determined to be only an optical companion we
do not provide coordinates for the companion.  Also, in general, we do
not include binary companions discovered since the publication of
GJ91; in almost all cases, these companions have accurate positions
reported in the literature.

\section{Positional Data for Gliese Stars}

The data available for stars that were added to the nearby stars catalogs varied greatly according to the exact pedigree of the star.  In
some cases, the positional data were of high quality (accuracy to a few arcseconds).   In many other cases, however, the positions were poorly known.  
Often, this was captured in the catalog by only listing the positions
to the nearest arcminute or by not listing any proper motion. In GJ91, 455 stars have their declination listed to arcminute
precision; another 48 stars have no listed proper motion.  All of the other RA's in GJ91 are listed only to integer seconds of time precision.
Many of the stars that have their declinations listed to 0.1 arcminute precision actually have positions known to much lower accuracy.

\subsection{Automated Recovery of Positions and Validation}

About 60\% of the Gliese stars are easily cross-matched to stars in the Hipparcos catalog based
either on a close coordinate match or on aliases to other catalog names provided in the Gliese
catalogs and the Hipparcos Input catalog.  These stars therefore 
have accurate J2000 positions and proper motions as determined by Hipparcos, and
with only a couple exceptions (which we discuss in the Appendix) we 
have simply adopted the Hipparcos astrometry.  
To aid in additional identifications, we used those epoch 1991.25 positions and proper motions to predict positions at 2MASS epoch 
and determined the 2MASS counterparts for each Hipparcos star.   

A few Gliese stars were present in the Hipparcos input catalog (HIC), but did not yield good 
fits from the Hipparcos mission data and therefore do not have Hipparcos (output) catalog 
data.   We used the data in the HIC and Gliese catalogs and DSS and 2MASS finding charts 
to identify these stars and determine their 2MASS cross-identifications and J2000, 
epoch 2000 positions.   A few additional stars with very poor positional data in the
Gliese or Woolley catalogs were discovered by us to be associated with Hipparcos stars during
our detective work.  Where warranted, we have included a discussion of these stars in
the Appendix.

For the 1555 Gliese stars without Hipparcos positions, we adopted an automated 
process for finding J2000 positions.  
To start the process, we used the 1950 positions and proper motions to predict 
J2000, epoch 2000 positions.  We then used the IRSA Gator 
service\footnote{http://irsa.ipac.caltech.edu}
to identify all 2MASS stars within two arcminutes of these positions.  We 
then queried the Bakos et al.\ (2002), Salim and Gould (2003), and Lepine 
et al.\ (2005) catalogs of high proper motion stars for position matches 
to the 2MASS stars.  For all position matches, we compared aliases in the 
high proper motion catalogs to those in the Gliese catalogs.  Where the 
aliases indicated we had found the correct 2MASS counterpart to the Gliese 
star, we also made sure that the optical - near-IR colors were reasonable.  
For binary Gliese stars, we insured that the correct binary component 
was matched to the correct 2MASS source.  Accurate J2000 positions for 
$\sim$800 stars were derived with the automated process.

The remaining approximately 700 Gliese stars were handled manually.  Finding 
charts from the first and second epoch DSS and from 2MASS were examined 
carefully.  In most cases, the Gliese stars have high enough proper 
motions and distinctive enough colors that it was possible to identify the star
unambiguously from these charts.  The IRSA Gator service was again utilized to 
determine the name, J2000 (2MASS epoch) position and $JHK_s$ photometry for 
these Gliese stars.  Where an unambiguous match was not apparent, we consulted 
papers referencing the star (as listed by SIMBAD) for additional finding charts or other 
identification clues.  For stars with Gliese position errors greater than 30$\arcsec$
and where serious detective work was required, we provide a brief 
synopsis of our process in the Appendix.  

Figure \ref{finderCharts} illustrates some of the difficulties posed
in identifying the positions for three individual systems.
The Hipparcos coordinates for GJ 563.2 AB, for example, are unreliable
(panel A).
In other cases, proximity of two similar brightness 
(but unassociated) stars has led to misidentifications in the 
literature (GJ 4327; panel B) or will lead to confusion in the future
due to the proper motion of the Gliese star causing it to pass 
very near the line of sight to another star
(GJ 3999 A; panel C).

Table \ref{mainTable} provides the J2000 positions 
and 2MASS data for the 4251 stars for which we were successful
in identifying their positions.
(Note that there are only 4106 lines of data, since some binary pairs
have fully blended positional/photometric information.)
Most positions are from Hipparcos (2693 stars) or from 2MASS (1549);
only 9 stars have positions from other catalogs,
as listed in Table \ref{mainTable}.
Table \ref{nonstars} provides 
information for the remaining 11 objects where we either could not determine a 
position or believe the object is non-stellar.

\subsection{Explanation to Table \ref{mainTable}}

As described in the introduction, the nearby star catalogs have evolved 
as new nearby stars were identified and 
as original members were dropped based on new estimates of their distances 
placing them outside the adopted distance limit.  
In Table \ref{mainTable} we have chosen to include nearly all
stars that appear in any of the nearby star catalogs -
Gl69, GJ79, GJ91, and Wo70.
That is, a star that was present in the 1969
Gliese catalog but not in the 1991 Gliese-Jahreiss catalog will still be
present in our Table \ref{mainTable}.   
Some distant giants ($>$100 pc; see Table \ref{outliers}) and 
Pleiades cluster members ($\sim$133 pc; ``g'' flags in Table \ref{mainTable})
remain in the table in spite of their nominally large distances.
Only 11 objects previously identified as nearby stars
are excluded from Table \ref{mainTable}:
2 sources that are actually QSOs,
2 proposed companions to brighter stars which we believe do not exist,
4 stars that are only optical companions,
1 probable plate flaw, 
and 2 stars which simply remain un-recovered.  
These misidentified objects are listed in Table \ref{nonstars}.
The columns for Table \ref{mainTable} are described here:

\medskip
\noindent{ {\bf Column 1:} Star Name}

The prefix for each star name is an indicator of the original source catalog:
Gl = from the Gliese (1969) catalog;
GJ = not in Gl69 but present in GJ79 or GJ91;
Wo = not in any of Gl69, GJ79 or GJ91, but in Woolley et al.\ (1970).
In some cases binary component letters have been added to the original names,
based on companion identification subsequent to GJ91.
Such cases are marked by a ``d'' flag in the comments (column 12).

\noindent{ {\bf Column 2:} Alternative Name}

Where a given star has two nearby star catalog names, the second name appears here.
If there is no nearby star catalog name and a Hipparcos name is
available, it is provided here.

\noindent{ {\bf Columns 3 and 4:} Positions}

Right ascensions and declinations are given for equinox and epoch 2000.  
For all but a few stars, the positions are based on Hipparcos or 2MASS data 
(as noted in column 12).  
For most binary stars with separations of five arcseconds or less, we list the binary pair on one row of the table with an AB designation.  
For binaries with Hipparcos positions for each component and separations $<$ 5$\arcsec$,
each star has its own row in the table, but we generally associate both
components to the same (unresolved) 2MASS object.

\noindent{ {\bf Columns 5 and 6:} Proper Motions}

Column 5 and 6 list the proper motions in the RA and Dec directions in $\arcsec$/yr.
The proper motions generally come from Tycho-2, Hipparcos, or GJ91,
in that order of precedence.
For the very small number of stars lacking reliable proper motions
in any of these three catalogs,
a handful of proper motion measurements from 
the USNO-B1.0 (Monet et al.\ 2003) and
UCAC3 (Zacharias et al.\ 2010)
catalogs have also been included.
For some binaries, the proper motion of the individual components
has been compromised by their proximity,
particularly for Hipparcos measurements.
To correct for this, we have enforced the alignment of
proper motion vectors for binary components within 10\arcsec\ of each other.

\noindent{ {\bf Columns 7 - 10:} 2MASS Name \& Photometry}

These columns list the 2MASS designation associated for each star and 
the corresponding J (1.2 $\micron$), H (1.6~$\micron$), K$_s$ (2.2 $\micron$) 
2MASS photometry.
Those binary companions with separations $\lapp$ 5$\arcsec$ have the same photometry listed for each component. 
Where the 2MASS photometry for a binary star is likely to be blended,
we indicate this with a ``f'' flag in the comments column.

\noindent{ {\bf Column 11:} Comments}

Comment flags listed here are explained in the Table \ref{mainTable} footnotes.

\noindent{ {\bf Column 12:} Coordinate Source}

This column notes whether the position of the star originates from the Hipparcos or 2MASS catalogs or an
additional source such as the LSPM catalog (Lepine et al.\ 2005).

\section{Properties of the Complete GJ Sample}

\subsection{Positional Accuracy}

We assert that we have determined arcsecond accuracy coordinates for all 
but a handful of the Gliese stars.  
Compared to the original source catalogs,
SIMBAD provides improved positions for many Gliese stars
and indeed 78\% of our positions agree with SIMBAD 
(as of April 2010)
to within an arcsecond.
Nevertheless there are still many Gliese stars whose positional listing in
SIMBAD is significantly in error.   
As a symptom of the less well known positions, 
SIMBAD only lists 2MASS cross-ids for 3293 of the 4251 Gliese stars,
whereas we report 2MASS ids and photometry for all but 14.
The number of stars whose positions in Table \ref{mainTable}
differ from SIMBAD 
by $>$ 10$\arcsec$ is 315, with 52 of these different by more 
than an arcminute.
An additional 85 stars from Table~\ref{mainTable} are either not listed
in SIMBAD or do not have coordinates there.
For the Gliese stars listed in SIMBAD, Figure \ref{simbadfig} shows 
the difference between our coordinates and those provided by SIMBAD.   
 
\subsection{2MASS Photometry}

For the 2MASS photometry presented in Table \ref{mainTable}, 
Figure \ref{2massColors} compares the $H-K_s$ and $J-H$ colors.
Most stars follow a well-defined color sequence
extending from white dwarfs in the lower left up into 
the main-sequence which follows an L-shaped central locus of point.
The densest pack of stars corresponds to main-sequence M stars,
with the very latest-type stars trailing to the upper right.
Outliers (shown as red points in Figure \ref{2massColors})
are detailed in Table \ref{outliers}.

\section{Conclusions} 

Because they are nearby and hence relatively bright, Gliese catalog stars often serve as template objects.  
In particular, nearby stars are among the most important targets in the sky
for surveys of planets and planetary debris 
(e.g.\ Butler et al.\ 2006; Bryden et al.\ 2006).
Their attractiveness as survey targets comes not only from their relative 
brightness but
also because they subtend the greatest angles 
for high resolution imaging of their disks and, potentially, planets.
Not surprisingly,
many of the most exciting planetary discoveries have been for Gliese stars,
including
the first detected debris disk
(Vega/GJ~721; Aumann 1984),
the first resolved disk
($\beta$ Pic/GJ 219; Smith \& Terrile 1984),
the first planet detected around a main-sequence star
(51~Peg/GJ~882; Mayor et al.\ 1995),
the first multiple-planet system
($\upsilon$~And/GJ 61; Butler et al.\ 1999),
and the first imaged planet
(Fomalhaut/GJ 881; Kalas et al.\ 2008).
Future surveys for planets via astrometric wobble (Unwin et al.\ 2008), infrared radial-velocity variation (Bean et al. 2009),
or direct imaging (Beichman et al.\ 2007)
will similarly focus on our closest neighbors.
It is therefore important that the properties of these stars be 
tabulated in an accurate and homogeneous fashion.  

While the quality of astronomical observations is normally limited 
foremost by the brightness of the targets, 
the proximity of the stars in the Gliese catalogs 
results in unique limitations - 
their high proper motion and large number of well-separated binaries
create serious difficulties in proper identification of the stellar
positions.
The net result is 
a surprisingly high degree of inaccuracy in the positions of
stars that would otherwise be expected to be the among the 
best in the sky.
We have taken a first step towards providing better data for the Gliese stars
by providing accurate J2000 positions and $JHK_s$ photometry for nearly all Gliese catalog stars.  
In Paper II of this series, we will provide published BVRI photometry for a large fraction of these stars.

\acknowledgments {We would like to acknowledge a careful reading 
of the manuscript by an anonymous referee.
This publication makes use of data products from 
the NASA/IPAC/NExScI Star \& Exoplanet Database (NStED),
the Two-Micron All Sky Survey (2MASS), 
the NASA/IPAC Infrared Science Archive (IRSA), and
the SIMBAD and VIZIER databases operated at CDS Strasbourg.
Some of the research
described in this publication was carried out at the Jet Propulsion
Laboratory, California Institute of Technology, under a 
contract with the National Aeronautics and Space Administration.}

\begin{appendix} 
\medskip

\noindent{\bf Gl 63} - Examination of DSS and 2MASS images shows two stars near the GJ91 position at
  the epoch of the 2MASS images.  GJ 63 is the westernmost, high proper
  motion component of this optical binary, and is 2MASS J01382162+5713571.

\medskip

\noindent {\bf Gl 127.1 B} -  Listed as an AB pair in the GJ91 catalog, with a
separation of 8$\arcsec$, but no position angle indicated.  GJ 127.1 A is a
well-known star with a Hipparcos position, so there is no difficulty
identifying A.  The Gliese catalog lists 
$V$=11.43, $B-V$=0.01 for A, $V$=14.73, $B-V$=0.62 for B.  Caballero
and Solano (2007), however, state that the ``status of the hypothetical
companion B is unknown''.  We examined all available on-line images of
the field (including 2MASS, where the A-B contrast should be best), 
but we find no sign of a B
component, at any position angle, with a separation of 8$\arcsec$ from A.  We
are therefore skeptical that B exists.

\medskip

\noindent{\bf Gl 130} - Examination of DSS and 2MASS finding charts shows a star
of the right magnitude, color and proper motion within a few
arcseconds of the position indicated in the GJ91 catalog.  This
star is 2MASS J03122972-3805204.  We believe this is the correct
ID.   This disagrees with the position listed in SIMBAD (by
about 30$\arcsec$), which is derived from Bakos (2002).  Our choice agrees
with the position reported by Rousseau \& Perie (1997).

\medskip

\noindent{\bf Gl 293} - Examination of the 2MASS and DSS finding charts shows a star
of the right proper motion at $\sim$30$\arcsec$ from the position indicated in the GJ91 catalog. 
This star is 2MASS J07530814-6747314.

\medskip

\noindent{\bf Gl 451 B} - Discovered by van de Kamp in 1968 as an apparent flare star 
companion, separated by 2$\arcsec$ from A.  Not detected since 1968 to our 
knowledge (e.g.\ Heintz 1984).  

\medskip

\noindent{\bf Gl 452 B} - Identified as a faint, blue companion about 8$\arcsec$ to east of 
Gl 452 A by Luyten.  We see a star of the right magnitude and location 
in the 1954 blue DSS plate.  However, it is not common proper motion 
with Gl 452 A.  Reid \& Gizis (2005) report they could not detect B at 
the expected position with the Keck I guide TV (at epoch $\sim$2002), with an estimated 
limiting magnitude of R $\sim$21.  We suspect Gl 452 B does not exist, at least
as a physical companion to A.

\medskip

\noindent{\bf Gl 534.1 B} - Location of B relative to A given incorrectly in GJ91.  
Correct relative position given in the CCDM.  Identification of B and 
confirmation of common proper motion with A made by examination of DSS 
and 2MASS finding charts.

\medskip

\noindent{\bf Gl 549 C} -  Gl 549 was listed as a triple system in Gl69, with the C 
component being located almost 1.5 degrees north of the AB pair.  In 
GJ91, the C component was no longer listed as a member of the system.  
C is real, and does have a proper motion quite similar to the AB 
pair.  However, it has a parallax which is different from the AB pair 
(Gliese \& Jahreiss 1988), and its spectral type and 
apparent magnitude are also inconsistent with forming a physical 
system with AB.

\medskip

\begin{figure}
\begin{center}
\includegraphics[width=4.7in]{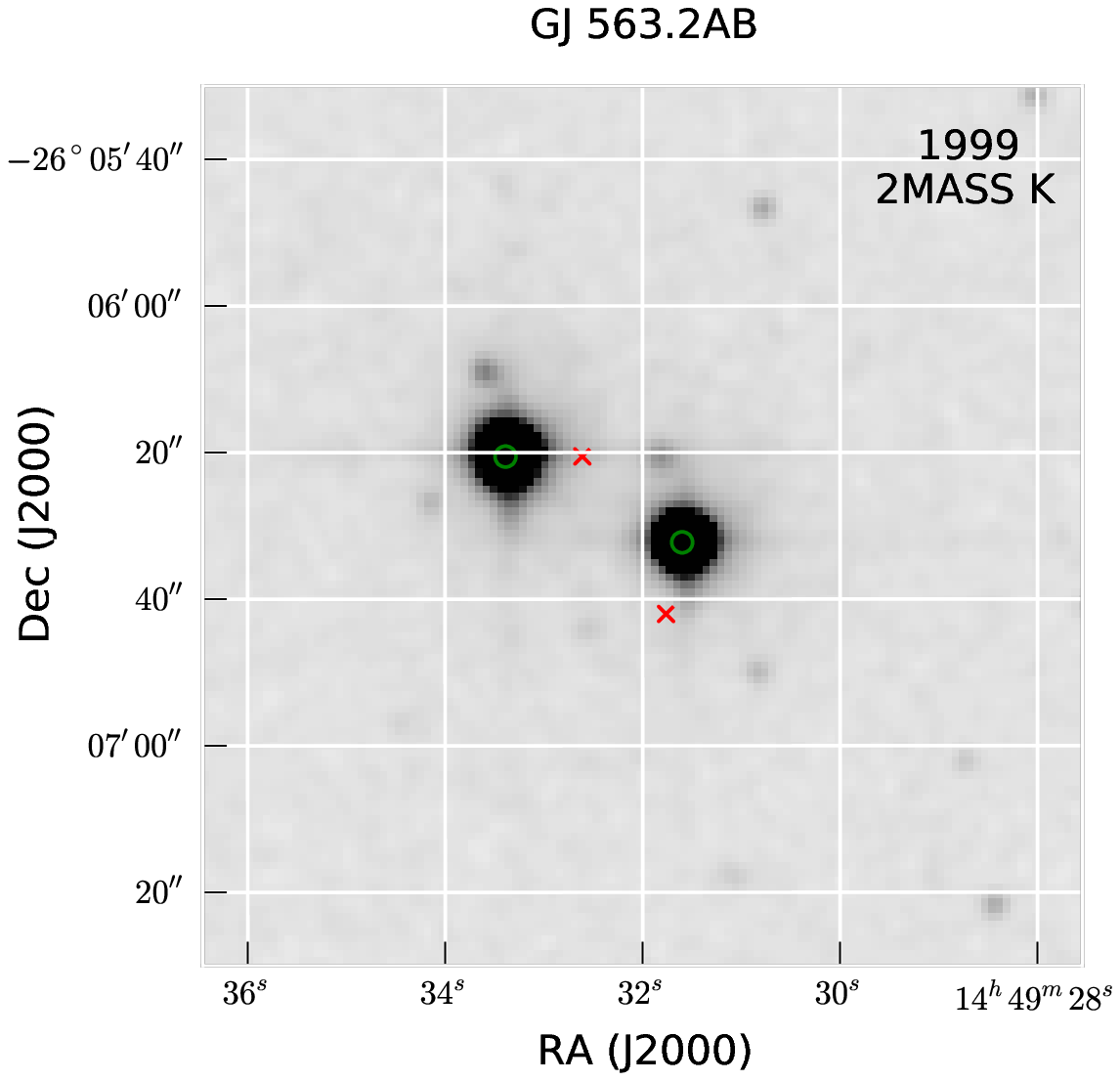} 
\end{center}
\caption{Illustrative images of three fields.
In each case, a 2$\arcmin$ by 2$\arcmin$ region is shown around
the target(s) of interest.  The stars' J2000 positions (from Table \ref{mainTable})
are shown as green circles.
Panel A shows the 2MASS K band image for GJ 563.2 AB.
The corresponding Hipparcos positions (projected to epoch 2000; shown as red crosses)
are clearly inconsistent.
In Panel B, the 1953 and 1995 DSS images show GJ 4327 approaching a non-moving field star
HIP 114994 (indicated by a red cross), causing confusion in their identification.
Panel C shows GJ 3999 A/4000 B at three epochs - 1954, 1989, and 1999.
As the pair moves southward, GJ 3999 A 
will soon become nearly coincident with the non-moving field star.
See the appendix for additional details on each of these systems.
}
\label{finderCharts}
\end{figure}

\noindent{\bf Gl 563.2 AB} - Alternate names for this binary are HIP 72509
and HIP 72511, and LHS 379 and LHS 380.  Normally we would have just accepted the Hipparcos
positions.  However, the J2000, epoch 2000 positions derived from the Hipparcos
data are clearly inconsistent with what is seen in the DSS and 2MASS charts (see Figure \ref{finderCharts}).
The Hipparcos positions indicate that the two stars are separated by 11$\arcsec$
in RA and 21.5$\arcsec$ in DEC.  However, visual examination of the finding charts
clearly indicates the stars have separations more like 20-25$\arcsec$ in RA and
of order 10$\arcsec$ in DEC.  We have chosen to adopt the 2MASS positions for
the system, and note that these positions agree with what is given in the HST
GSC 2.2 and by the Carlsberg Meridian Catalog, while they disagree by 10-15$\arcsec$
with the Hipparcos positions.

\medskip

\noindent{\bf Gl 601 B} - The notes to G69 identify this star with L 153-157, with a 
position offset from the A component of 157$\arcsec$ at PA=258$^\circ$.  A star with 
the expected motion is visible near the predicted position.  The only 
astrometric catalog we can find for this star is the UCAC3, which lists 
15 54 43.49, -63 26 14.8 (epoch 1998.3).

\medskip

\noindent{\bf Gl 629.2 B} - The notes section of G69 indicate that this is a common proper 
motion companion to Gl 629.2 A. Giclas (1960) provides 1950 coordinates and 
a finding chart for Gl 629.2 B under the designation G17-27.  That finding 
chart allows easy identification of this star as 2MASS J16352902-0357571.

\medskip

\noindent{\bf Gl 732 B} -  This star probably does not exist.  Separation from A listed 
as about 10$\arcsec$ and about 3.3 magnitudes fainter.  Proper motion given as 
about 1.0$\arcsec$/yr to PA=161$^\circ$.  Examination of the DSS and 2MASS charts 
shows no evidence for a companion to A.  We could find no 
post-discovery evidence in the literature that B exists.

\medskip

\noindent{\bf Gl 734 AB}  - The GJ91 catalog indicates that B is separated by 5.2$\arcsec$ at PA=21$^\circ$ from 
A, with B being about 2.5 mag fainter than A in the visual. Similar 
information is listed in the WDS and CCDM. Visual inspection of the 
DSS and 2MASS images allows easy identification of A, but no strong 
evidence for the presence of B. The 1952 DSS images show a hint of 
extension in the NS direction, with a hint of a second set of 
diffraction spikes northward of the main spikes, as expected. The 
2MASS PSC shows two sources when queried, the brighter being A and the 
fainter displaced from A by about 4$\arcsec$ to the NNE. This second source is 
barely detected by 2MASS, but is at the approximate position expected for B. 
We adopt the 2MASS coordinates; a better measurement would be useful.

\medskip

\noindent{\bf Gl 762 } -  Nothing present within 30$\arcsec$ of the GJ91 coordinate.  
Examination of the DSS and 2MASS charts shows a star of the right 
(relatively large) proper motion and colors about 45$\arcsec$ from the 
catalogued position.  No other star of similar magnitude is within 2.5 
arcminutes.

\medskip

\noindent{\bf Gl 774 AB } - The Hipparcos input catalog lists GJ 774 A as an
alias to HIC 98811 A, providing a J2000 coordinate of 20 04 02.1, -65 35 58.1.
However, that is approximately the correct coordinate for GJ 774 B (and
about 20$\arcsec$ from its true position).  HIP 98811 A should instead be aliased
to GJ 774 B.  The GJ91 coordinate and proper motion allow the J2000 position
to be estimated to better than 30$\arcsec$ accuracy, and the proper motion is large
enough to be distinctive - thus allowing easy identification of the correct star.

\medskip

\noindent{\bf Gl 781.1 AB} - The GJ91 catalog provides individual RA/Dec values 
for both components, with the secondary indicated as being 3 seconds west and 
18$\arcsec$ north of the primary.  When the GJ91 1950 coordinates are 
precessed and proper motion corrected to J2000, epoch 2000 and those 
positions checked versus the 2MASS images, there are two relatively 
bright stars within a few arcseconds of where they are predicted to be 
- A: 20 07 48.0, -31 45 30.  (2000) and B: 20 07 45.0, -31 45 14.  
(2000).  Comparing the 2MASS and DSS images shows those stars to both 
have the proper motions expected (0.8$\arcsec$/yr towards PA 159).  The only 
problem is that Gl 781.1 A was included in the Hipparcos input catalog 
as HIC 99150.  A star was found by Hipparcos at almost exactly the 
predicted HIC location, and subsequently HIP 99150 was given the 
Hipparcos 1997 position of 20 07 44.98, -31 45 14.4 (2000).  
Unfortunately, based on the data in GJ91, this star is not Gl 781.1 A 
but is instead Gl 781.1 B.  SIMBAD adopted the Hipparcos identification, 
and therefore called the star at 20 07 48.0, -31 45 30 Gl 781.1 B.  We 
prefer to retain the positional reference data from the GJ catalog 
itself - making the A component the south-eastern star of the binary.
The two stars have very similar luminosities.

\medskip

\noindent{\bf Gl 806.1 B} - The A component of this system is a $V$=2.45 K giant. 
According to GJ91, the B component is located 6s west and 6$\arcsec$ south 
from A, and is listed as a $V$=13.4, M4 dwarf. We could not confirm B by 
its proper motion using the DSS and 2MASS charts because the image of 
A is so bright. However, the 2MASS images show two stars at about the 
expected position for B. Neither star is in the on-line 2MASS PSC. However, 
both stars are in the 2MASS ``working database''. One of this pair is 
too blue to be an M star. The brighter and more NW of this pair does 
have near-IR colors compatible with being an M4 dwarf. That star is at 
20 46 06.39 +33 58 06.1 (2000), observed by 2MASS on 16 May 1999, with 
$J$=9.456, $H$=8.889,$K$=8.649 (uncertainties of 0.031, 0.015 and 0.009) - data
from the 2MASS working database.

\medskip

\noindent{\bf Gl 871.1 AB}  - This binary appears in the Gl69 catalog, but not 
GJ91. Spectral types are listed as M3 and M4. The positions in 
Gl69 were only given to a precision of 0.1 minute of time, and 1$\arcmin$ of 
declination; the pair has negligible proper motion. No star appears 
within 1$\arcmin$ of the stated position on the DSS and 2MASS charts, but the 
binary is clearly identifiable on those charts (based on relative 
position and colors) about 2$\arcmin$ west of the catalogued position.

\medskip

\noindent{\bf GJ 1001 AB} - GJ 1001 was listed as a single star in the GJ79 update 
to the Gliese catalog. Using the coordinates from GJ79, no compatible 
object is within 1$\arcmin$ on the DSS or 2MASS charts. However, a star with 
the right proper motion (1.6$\arcsec$/yr towards PA=155$^\circ$) is visible about 2$\arcmin$
south of the expected position. A common proper motion companion is 
also visible in those charts to the west of the primary. This pair is 
now referenced as GJ 1001 A and 1001 B (see Goldman et al. 1999). 
Both stars have 2MASS counterparts, which we have adopted for our 
position reference.

\medskip

\noindent{\bf GJ 1022} - Examination of DSS and 2MASS charts showed no star within 1$\arcmin$ 
of the expected position from GJ91. However, a star matching the 
catalogued proper motion (1.119$\arcsec$/yr to PA=94$^\circ$) is easily identifiable on 
those charts about 1.5$\arcmin$ from the GJ91 position. The 2MASS colors of 
that star  (2MASS J00492903-6102326) are compatible with the late-M 
spectral type given for this star by GJ91.

\medskip

\noindent{\bf GJ 1167 B} - GJ 1167 A is easily identified in DSS and 2MASS charts, within 
15$\arcsec$ of the position given by GJ91. GJ91 states that ``comp B doesn't 
exist''. However, in the GJ79, a GJ 1167 B is listed (at a position about 
80$\arcsec$ east and 3$\arcmin$ north of the A position). A star is present in the DSS 
and 2MASS charts within about 15$\arcsec$ of the GJ79 prediction, and with 
proper motion between the 1955 epoch DSS and 2000 epoch 2MASS charts 
essentially identical (to the eye) as the A component. The photometry 
for B is significantly bluer than for A, however, which seems 
unphysical since A is 2-4 mag brighter than B. This suggests the pair 
is not a physical binary, in agreement with the statement in GJ91.

\medskip

\noindent{\bf GJ 1287} - Despite a true position more than 30$\arcsec$ from what is given in 
GJ79, this star is easily identified on DSS and 2MASS charts due to 
its large proper motion (0.9$\arcsec$/yr to PA=98$^\circ$).  It was not included in 
GJ91 because newer distance estimates place it beyond 25 pc.

\medskip

\noindent{\bf GJ 2013} - Examination of DSS charts shows no good candidate within one 
arcminute of the GJ79 position.  We identified the star using the 
finding chart in Eggen (1968).   This star is not included in GJ91 
because the revised distance estimate places it beyond 25 pc.

\medskip

\noindent{\bf GJ 2014} -  An accurate position for this star is provided by Costa et al 
(2006).  The corresponding 2MASS object is 2MASS J00495863-2624055.  
The near-IR and V-K colors agree with that expected for the $B-V$ color 
listed in GJ79.  This star was not included in GJ91.

\medskip

\noindent{\bf GJ 2023} - This is a very faint star with quite small proper motion.  It 
is therefore not possible to identify it unambiguously using the DSS 
and 2MASS charts.  Friedrich et al. (2000) were able to identify it and 
obtained spectra confirming it as a white dwarf.  Using the 
coordinates provided in that paper, we identify GJ 2023 as 2MASS 
J01295612-3055098.

\medskip

\noindent{\bf GJ 2052} - This is a faint, bluish star with little proper motion. Only 
one star is present within 30$\arcsec$ of the predicted position on the DSS 
charts. That star is not present in the 2MASS PSC, because it is too 
faint. A blue star with nearly the same V magnitude as listed in GJ79 
and located within 5$\arcsec$ of the GJ79 position is present in the Lick 
Northern Proper Motion Survey (Klemola et al. 1987). We adopt their 
position.

\medskip

\noindent{\bf GJ 2087} - There is no star in the DSS or 2MASS charts
at the J2000 GJ91 position and Bakos et al. (2002) 
claim this is a plate flaw.

\medskip

\noindent{\bf GJ 2133} - This star is located in a very crowded field.  No object with 
the expected proper motion and color is present on the DSS and 2MASS 
charts within one arcminute of the catalogued position.  A star with 
the expected characteristics is present on those charts about one 
arcminute west and 30$\arcsec$ south of the catalogued position.  That star is 
2MASS J17540302-3440177.

\medskip

\noindent{\bf GJ 2154 AB} - Examination of the DSS charts showed no stars 
matching the expected properties within one arcminute of the catalogued 
position.  However, the binary is easily identifiable about 1.5$\arcmin$ away based on the relatively large proper motion and the 
appropriate separation and position angle.

\medskip

\noindent{\bf GJ 3063} - This should be a relatively bright ($V$=11.9) M3
star of low proper motion according to the GJ91 catalog.  However, only two stars
appear in the 2MASS images near the catalogued position, and neither are particularly
red.  No star of approximately the right properties is present in, for example,
the USNO B1 or HST GSC within several arcminutes of the expected
position.  The only alias in GJ91 is to Sm177 (Smethells 1974, PhD thesis), and
SIMBAD provides no literature references.  We were unable to recover this star.

\medskip

\noindent{\bf GJ 3085} - No star of the expected properties is located on the DSS 
charts within one arcminute of the position listed in GJ91.  A finding 
chart is provided in Pesch \& Sanduleak (1978); based 
on that chart, we identify GJ 3085 as being located about 2.2$\arcmin$ north 
and 1$\arcmin$ east of the catalogued position. Its near-IR colors are 
consistent with the ``m'' spectral type provided in GJ91.

\medskip

\noindent{\bf GJ 3096} -  No star of the expected properties is located on the DSS 
charts within one arcminute of the position listed in GJ91.  A finding 
chart is provided in Pesch and Sanduleak (1978); based 
on that chart, we identify GJ 3096 as being located about 1.2$\arcmin$ north 
and 0.4$\arcmin$ west of the catalogued position. Its near-IR colors are 
consistent with the ``m'' spectral type provided in GJ91.

\medskip

\noindent{\bf GJ 3101} - No star of the expected properties is located on the DSS 
charts within one arcminute of the position listed in GJ91.  Reid, 
Hawley, \& Gizis (1995) list a position for GJ 3101 that 
is more than 5$\arcmin$ different from that in GJ91. Comparison of the DSS 
chart at the Reid et al. position to a finding chart in Pesch and 
Sanduleak (1978) allows identification of GJ 3101 (aka 
PS244). This star is 2MASS J01360872-2652161; the near-IR colors are 
consistent with the ``m'' spectral type listed in GJ91.

\medskip

\noindent{\bf GJ 3397} -  Inspection of the DSS images
show a star with the correct colors and proper motions about 0.8$\arcmin$ S and 0.1$\arcmin$ E of
the coordinate in the GJ91 catalog. This star is 2MASS J06322189-6957445. 

\medskip

\noindent{\bf GJ 3521} - Examination of the DSS charts shows no appropriate star at 
the catalogued position.  A star with the expected proper motion, 
magnitude and colors is present about one arcminute west and 10$\arcsec$ south 
of the GJ91 position.  This star is 2MASS J08550473-7135480.

\medskip

\noindent{\bf GJ  3592} -  We used the finding chart provided by Eggen (1969) to 
identify this star on the 2MASS charts.  This object is 2MASS 
J10145842-5611114.  This position agrees with that given by Wegner (1973).

\medskip

\noindent{\bf GJ 3618} - The very large proper motion of this star (1.5$\arcsec$/yr) makes 
its identification on DSS and 2MASS charts easy despite the more than 
one arcminute error in the catalogued position.

\medskip

\noindent{\bf GJ 3999 A / 4000 B} -  
This is a binary separated by $\sim$30$\arcsec$ at PA=338$^\circ$.  
Identification made difficult because GJ 3999 A is passing very near a 
much brighter field star at 2MASS epoch (see Figure \ref{finderCharts}).  
Its distinctive proper motion 
and colors make the identification secure.

\medskip

\noindent{\bf GJ 4012} - No star evident on the DSS and 2MASS charts within 2$\arcmin$ of the 
catalogued position.  We used a finding chart from Luyten (1949) - 
under the alias L270-37 - to identify the star on the DSS charts, and 
to confirm its proper motion by comparison to 2MASS charts.  Our 
2MASS-based J2000 position is within $\sim$5$\arcsec$ of the position listed by 
Downes et al. (2001).

\medskip

\noindent{\bf GJ 4028} - Catalogued position in error by $\sim$1.5$\arcmin$.  High proper 
motion makes identification on DSS and 2MASS charts secure.

\medskip

\noindent{\bf GJ 4078} -  There is a star 2.2$\arcmin$ E and 1$\arcmin$ S
of the GJ91 coordinate in the DSS images whose colors and proper motion
match those expected for GJ 4078. This star is 2MASS J18495119-5726486.
SIMBAD lists the star at this position as LHS 3413, one of the aliases
given for GJ 4078 in the GJ91 catalog.

\medskip

\noindent{\bf GJ 4088} -  In the GJ91 catalog, GJ 4088 is aliased to LP 571-80.  LP 571-80
is aliased to HIC 93047 in the Hipparcos Input Catalog.  However, the position
given in the HIC is considerably in error, and presumably as a result no
useful data were obtained by Hipparcos for this star.  The coordinates given
in the GJ91 catalog are in error by less than 30$\arcsec$, allowing easy identification of
the correct star based on its proper motion and magnitude.

\medskip

\noindent{\bf GJ 4103} - Very little information for this star is given in GJ91.
An alias is given to LP 336-6 (v.1 of the NLTT), and the listed
position and proper motion are taken directly from there (or
from the BPM catalog, where it is listed as BPM94000).
However, no object with the listed proper motion
is present within 2.5$\arcmin$ of the listed position.
A star of somewhat similar position and proper motion is
present several arcminutes north of the listed position, but
that star is identified in the NLTT as LP 336-7, and we
do not believe there is enough information to equate that
star to GJ 4103.  We do note that there is a 3rd NLTT star
somewhat nearby in space whose listed red and blue magnitudes
and proper motion size and direction are exactly equal to
those for LP 336-6 - this other object is LP 336-2.  A barely
resolved binary
of the appropriate general brightness and proper motion
is present on the DSS and 2MASS charts at about the
position listed for LP 336-2 - this pair is 2MASS J19050788+3237547 and
2MASS J19050756+3237526.  Weis (1987) also noted the
exact duplication of information in the NLTT for these two
LP stars and that no obvious candidate was present at the
listed position for LP 336-6.  Our conclusion (similar to
Weis), is that either
LP 366-6 does not exist, or it is probably the same object as LP 336-2.

\medskip

\noindent{\bf GJ 4112} - The DSS images show a bright star 30$\arcsec$ E and 8$\arcsec$ N of
the GJ91 coordinates whose proper motion and colors match those expected
for GJ 4112. This star is 2MASS J19340394-5225144. Pokorny et al. (2004)
lists the star at this position as having a proper motion consistent
with that given in the GJ91 catalog, and indicate this star is L275-26,
one of the aliases listed in the GJ91 catalog for GJ 4112.

\medskip

\noindent{\bf GJ 4133} - High proper motion star (1.08$\arcsec$/yr).
The DSS images have a bright star 1.1$\arcmin$ E and 45$\arcsec$ S of
the GJ coordinates whose proper motion and colors match those expected
for GJ 4133. This star is 2MASS J20053482-1056545. SIMBAD matches this
star to LHS 483 and LP 754-16, both aliases for GJ 4133 given in
the GJ91 catalog.

\medskip

\noindent{\bf GJ 4187 A / 4188 B} - True position slightly more than 30$\arcsec$ different from GJ91 coordinates, but binary easily identifiable on DSS and 2MASS charts.

\medskip

\noindent{\bf GJ 4189 A / 4190 B} - These two binary components have a separation of 4 
arcseconds in the GJ91 catalog and a PA of 341 degrees in the WDS 
catalog.  The stars are of a similar spectral type and are resolved in 
the DSS images where it is clear they share a common proper motion.  
Using the information in the WDS catalog we conclude that GJ 4190 B is 
the more northerly star.  This disagrees with the GJ91 information 
(which traces back to the NLTT), but agrees with the double star 
measures of Worley (1962).

\medskip

\noindent{\bf GJ 4191} - The GJ91 catalog indicates this star is a white dwarf
with a relatively high proper motion.  Examination of DSS images shows a blue star 
90$\arcsec$ E and 15$\arcsec$ N of the GJ position that has about the right
proper motion.  That star is 2MASS J21193652-5550144.

\medskip

\noindent{\bf GJ 4193} - DSS and 2MASS charts show no appropriate match within
one arcminute of the input coordinate.  There is a star of the right magnitude
and color about 20$\arcsec$ E and 1.6$\arcmin$ south of the GJ coordinate.  This star
is 2MASS J21261422-4227320.   Hawley, Gizis \& Reid (1996) give this as the
coordinate for BPM 43997, which is one of the aliases listed for GJ 4193 by
the GJ91 catalog.

\medskip

\noindent{\bf GJ 4224 AB} - No star is visible in the DSS or 2MASS charts within 1$\arcmin$ of the 
expected position based on GJ91 1950 position and proper motion. A 
star with the right magnitude, colors and proper motion (0.34$\arcsec$/yr to 
PA=121$^\circ$) is present about 1$\arcmin$ east and 1.5$\arcmin$ south of the GJ91 position. 
The GJ91 catalog does not provide PA or separation, but says the magnitude 
difference is zero. On the DSS charts, the star we have identified as 
GJ 4224 has faintly visible double diffraction spikes, indicating a 
binary separated approximately north-south by a few arcseconds. 
Weistrop (1980) indicates that GJ 4224 (under the alias Sm 
89) ``appears double at the telescope'', also indicating the separation 
is probably of order 2-3$\arcsec$. Based on its J2000 coordinate, the 
star we have identified as GJ 4224 is also HIP 107711.  We 
use the Hipparcos position for both components.

\medskip

\noindent{\bf GJ 4237 A / 4238 B} - Listed in the GJ91 catalog as being separated by 3.4$\arcsec$,
but no position angle given.  The 2MASS catalog lists two stars of about the right
magnitude and separation very near the GJ91 input position.  These two stars are
2MASS J21553860+3302138 and 2MASS J21553870+3302106.   The WDS indicates that B is
south of A, therefore we match GJ 4237 A to 2MASS J21553860+3302138, and GJ 4238 B
to 2MASS J21553870+3302106.

\medskip

\noindent{\bf GJ 4249 } - There is no star of appropriate magnitude within
two arcminutes of the GJ91 position.  A bright star with the appropriate
(large) proper motion is, however, easily found on the DSS and 2MASS charts
about three arcminutes east of the catalogued position.  Based on the J2000
position and magnitude, this star is HIP 108890.  We adopt the Hipparcos
position for GJ 4249.

\medskip

\noindent{\bf GJ 4327} - There are two essentially equal brightness stars at the position
indicated by GJ91. The westernmost of the pair is redder and has high 
proper motion to the south, as expected - we identify this star as GJ 4327. 
It is 2MASS J23172441+3812419.  GJ91 lists LHS 3293 as an alias.  
Unfortunately, the Hipparcos input catalog aliases LHS 3293 to HIP 114994, 
the easternmost (approximately stationary) star of the pair.  We disagree 
with this latter alias, and instead alias LHS 3293 to the 2MASS star indicated above.

\medskip

\noindent{\bf GJ 4386} - There is no appropriate star within 30$\arcsec$ of the GJ91 
coordinates; however, a star of the right magnitude, color and proper 
motion (based on examination of DSS and 2MASS finding charts) is 
present abut 60$\arcsec$S and 15$\arcsec$W of the original coordinate.  This star is 
2MASS J23594483-4404599.  SIMBAD cross-identifies this star with LTT 
9828 and LP 988-175, both aliases for GJ 4386 listed in GJ91.  The J-K 
color for this star seems much too blue for the M7 spectral type 
listed in GJ91.  The actual spectral type is probably much earlier - 
Reid et al.\ (1995) give the spectral type as M2.

\medskip

\noindent{\bf Wo 9354} -   The Woolley (1970) catalog indicates that this star is Vyss 606 (Vyssotsky 1956).
There is no star that matches the expected properties within 1$\arcmin$ of the input position.
Only one star of about the right magnitude and color is present within
about 2.5$\arcmin$ of the Woolley coordinate - it is 
2MASS J11163293+2627415.   Examination of the  Hipparcos 
input catalog indicates that this star is HIP 55077; the HIC provides
a cross-identification to Vyss 606, confirming the identification.  

\medskip

\noindent{\bf Wo 9358} -  No star of the appropriate properties falls within one 
arcminute of the Woolley catalog position.   A star of the correct (but relatively
small) proper motion and with the correct approximate magnitude lies about three
arcminutes south of the Woolley position.  This star is also 
HIP 55605.  
\medskip

\noindent{\bf Wo 9540 C} -  The A component of this system is $V$=4.2.  According to 
GJ91, the C component  is separated from A by 7.6$\arcsec$ at PA=53$^\circ$.  C is 
not present in the 2MASS PSC.  Because A and B are so bright, C can 
not be identified on the DSS or 2MASS charts.  However, a star at the 
right position relative to A is present in the HST GSC2.3.2, and we 
adopt the coordinate from that source for Wo 9540 C.

\medskip

\noindent{\bf Wo 9785} -  This star is the exciting source for the planetary nebula NGC 7293.
Use of IRSA's FINDER allows relatively secure identification of 
the star and a match to 2MASS J22293854-2050136.

\medskip

\noindent{\bf Wo 9805} - No star of the appropriate properties falls within two 
arcminutes of the Woolley catalog position.  However, the Woolley catalog
indicates that Wo 9805 is also Vyss 856, and the Hipparcos Input Catalog
cross-matches Vyss 856 to 
HIP 113782.  We confirm that the colors and (small) proper motion 
of the star at the Hipparcos position agree with expectations for 
Wo 9805 given its spectral type of M0.5 and BVRI photometry provided in 
Laing (1989).
\end{appendix} 

\setcounter{figure}{0}
\begin{figure}
\begin{center}
\includegraphics[width=4.7in]{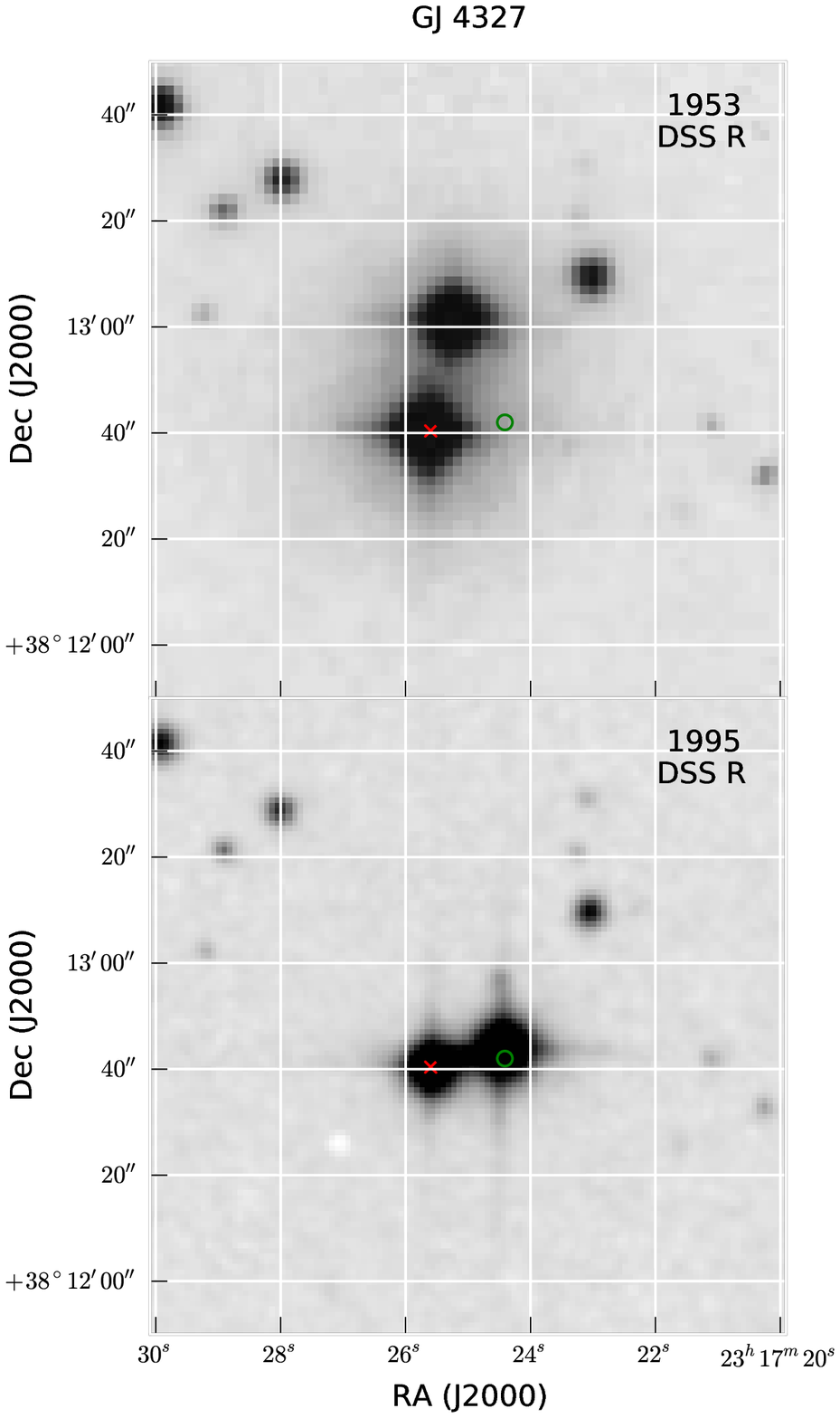}
\end{center}
\caption{
(Figure continued...)
}
\end{figure}

\setcounter{figure}{0}
\begin{figure}
\begin{center}
\includegraphics[width=4.7in]{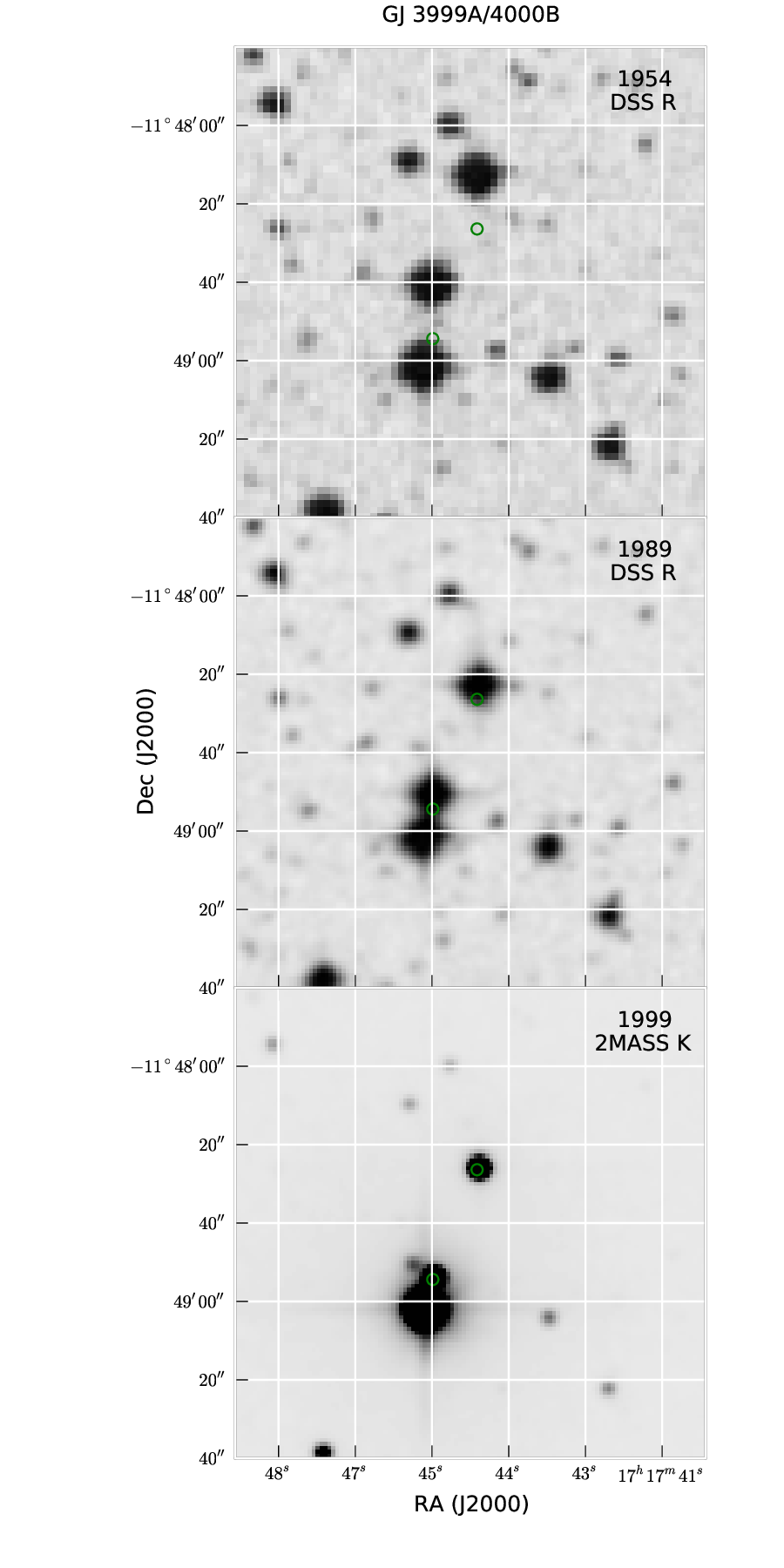}
\end{center}
\caption{
(Figure continued...)
}
\end{figure}

\begin{figure}
\begin{center}
\includegraphics[width=3.5in,angle=-90]{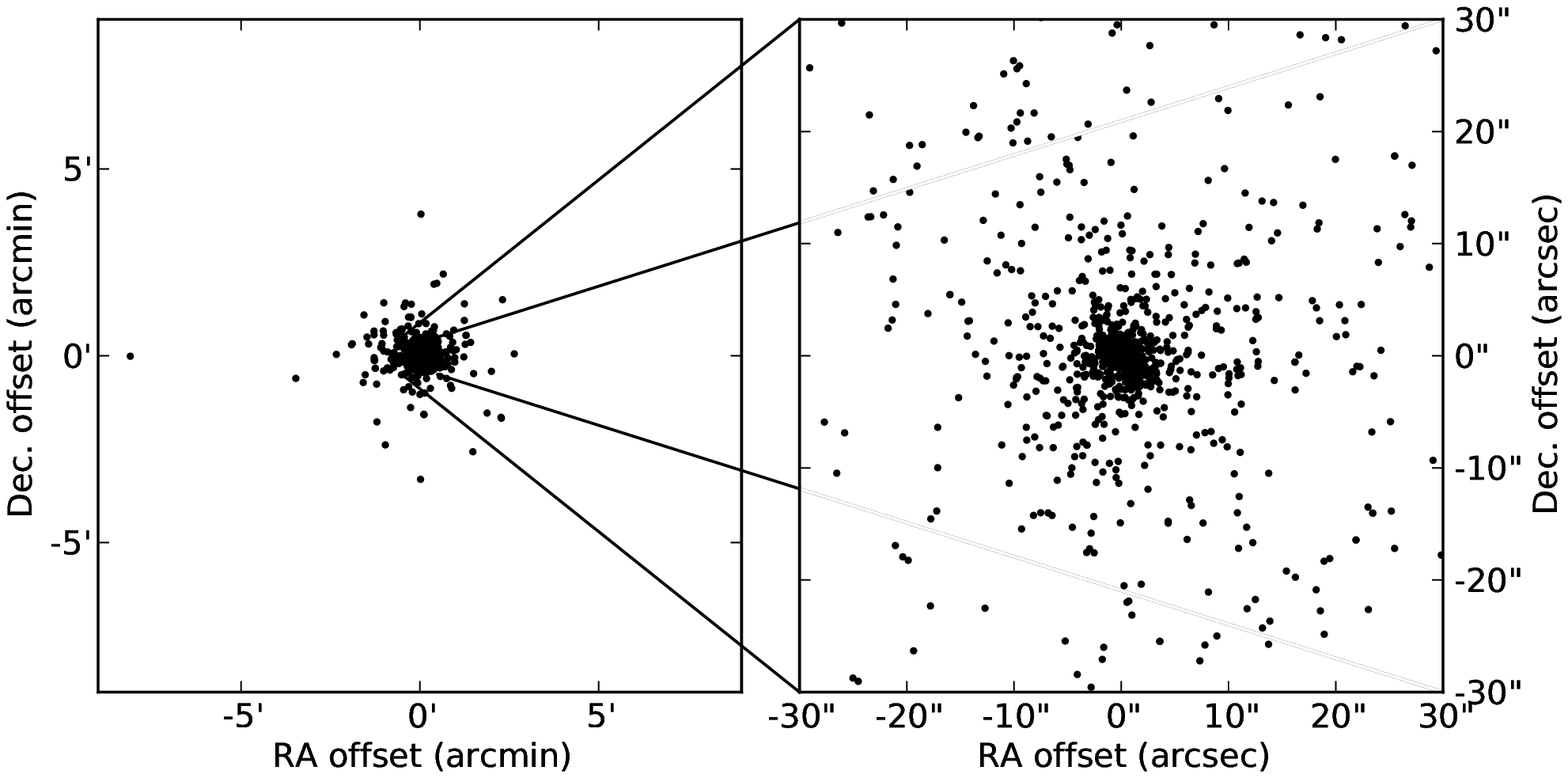}
\end{center}
\caption{For Gliese stars listed in SIMBAD (as of April 2010), 
we show the differences between our 
calculated positions and those given by SIMBAD.
The left panel displays stars with offsets of several arcminutes,
while the right panel zooms in to show differences on a finer scale.
The majority of the stars fall in the center of the plots,
with positions agreeing to better than 1 arcsecond.}
\label{simbadfig}
\end{figure}

\begin{figure}
\begin{center}
\includegraphics[width=5.7in]{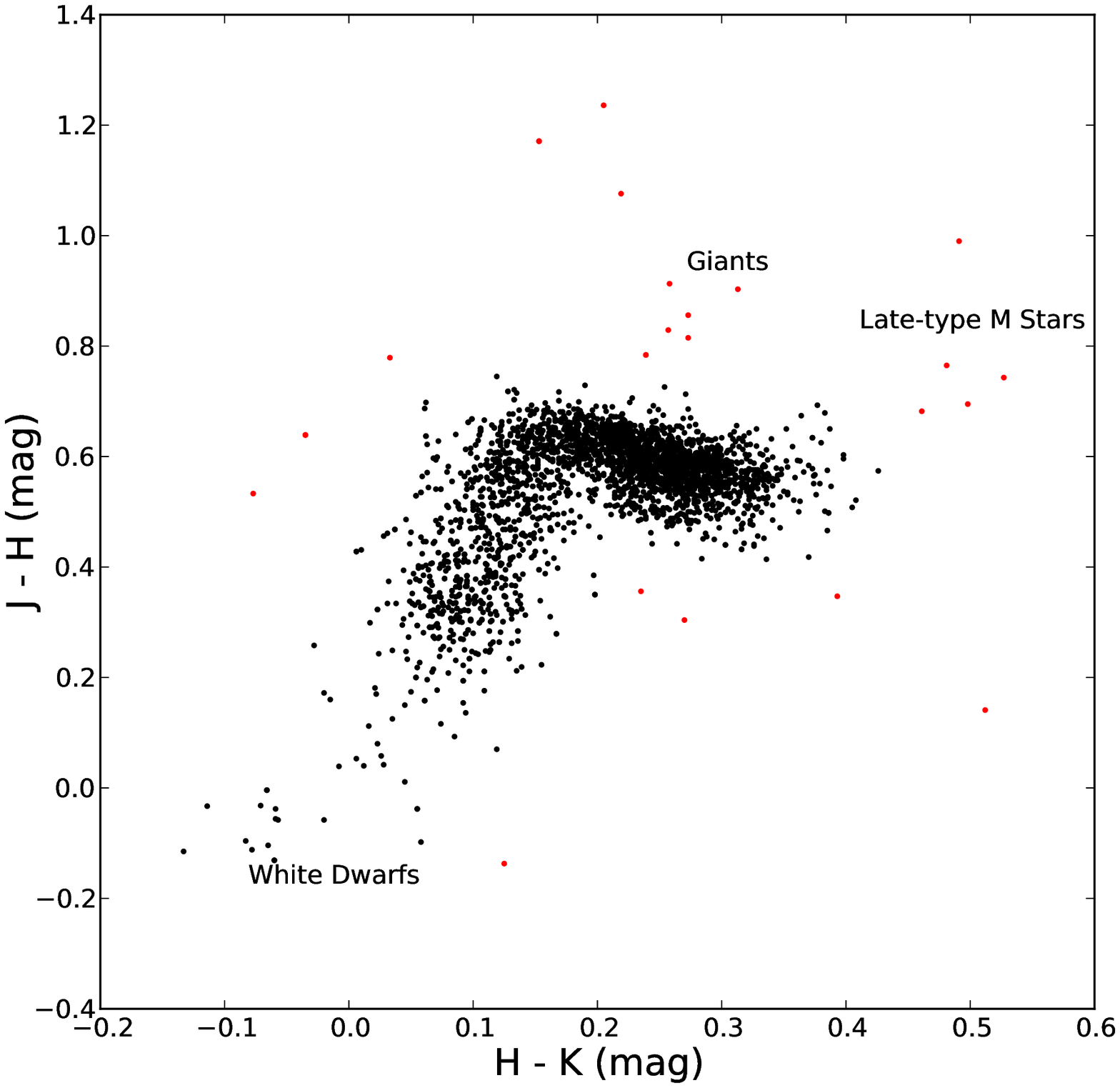}
\end{center}
\caption{Near-IR color-color diagram.
For the 2MASS photometry presented in Table \ref{mainTable}, $J-H$ is
plotted vs $H-K_s$.  Most stars follow a well-defined color sequence.
Outliers from the main distribution (red points)
are detailed in Table \ref{outliers}.}
\label{2massColors}
\end{figure}

\clearpage
\begin{deluxetable}{lcccrrcrrrcc}
\tablewidth{8.5in}
\rotate
\tabletypesize{\footnotesize}
\setlength{\tabcolsep}{0.05in}
\tablecaption{GJ Catalog \label{mainTable}}
\tablehead{				
\colhead{Name$^a$ }		&	\colhead{Alt name$^b$}	&	\colhead{RA}	&	\colhead{Dec}	&	\colhead{$\mu$ RA}	&	\colhead{$\mu$ Dec}	&	\multicolumn{4}{c}{2MASS}	&	\colhead{Comments}	&	\colhead{Coord. Source}	\\
		&		&	\colhead{J2000}	&	\colhead{J2000}	&	\colhead{\arcsec/yr}	&	\colhead{\arcsec/yr}	&\colhead{Name}		&	\colhead{$J$}	&	\colhead{$H$}	&	\colhead{$K_s$}	&		&											}
\startdata		
Wo  9846	&	HIP 18	&	00 00 12.68 	&	$-$04 03 13.3 	&		$-$0.123	&		0.024	&	J00001269$-$0403133	&	8.599	&	7.999	&	7.851	&c&	Hipparcos	\\
GJ 1293	&	Wo 9847	&	00 01 25.84 	&	$-$16 56 54.4 	&		0.298	&		$-$0.258	&	J00012581$-$1656541	&	8.017	&	7.408	&	7.217	&&	Hipparcos	\\
Wo  9848	&	HIP 149	&	00 01 54.68 	&	+26 00 14.6 	&		$-$0.310	&		$-$0.631	&	J00015470+2600153	&	9.372	&	8.859	&	8.743	&&	Hipparcos	\\
GJ 4388	&	HIP 150	&	00 01 55.27 	&	$-$37 13 47.8 	&		$-$0.020	&		$-$0.046	&	J00015527$-$3713479	&	5.186	&	4.665	&	4.447	&&	Hipparcos	\\
GJ 1294 A	&	HIP 169 A	&	00 02 08.73 	&	$-$68 16 50.8 	&		0.195	&		$-$0.210	&	J00020875$-$6816509	&	6.736	&	6.113	&	5.896	&&	Hipparcos	\\
GJ 1294 B	&	HIP 169 B	&	00 02 09.35 	&	$-$68 16 53.2 	& 0.195 & $-$0.210 &	J00020939$-$6816534	&		&	6.629	&		&&	Hipparcos	\\
Gl  914  AB	&	HIP 171	&	00 02 10.15 	&	+27 04 56.1 	&		0.830	&		$-$0.989	&	J00021014+2704570	&	4.702	&	4.179	&	4.068	&e,f&	Hipparcos	\\
Gl  915	&		&	00 02 10.73 	&	$-$43 09 55.6 	&		0.616	&		$-$0.677	&	J00021076$-$4309560	&	12.597	&	12.425	&	12.445	&&	2MASS	\\
GJ 3001	&		&	00 02 40.10 	&	$-$34 13 39.5 	&		0.150	&		$-$0.743	&	J00024008$-$3413386	&	14.117	&	14.024	&	13.919	&&	2MASS	\\
... \\
... \\
\enddata	
\tablenotetext{a}{Gl = from the Gliese (1969) catalog;
  GJ = not in Gl69 but present in GJ79 or GJ91; 
  Wo = not in any of Gl69, GJ79 or GJ91, but in Woolley et al.\ (1970).}
\tablenotetext{b}{For stars with both Gliese and Woolley names, 
we give the Wo name here (and the Gl name in Column 1).}
\tablenotetext{c}{Star is not in the GJ91 catalog.}	
\tablenotetext{d}
  {Not listed as a member of a binary system in Gl69, GJ or Wo catalogs; 
  binary companion identified subsequent to GJ91.}
\tablenotetext{e}{The secondary star is separated from the primary by 
  less than 5\arcsec\ and HIP/2MASS positions are not available for both stars,
  so we list them as one line in this table.}
\tablenotetext{f}{GJ binary where 2MASS photometry is probably blended, most separations $<$5\arcsec.}
\tablenotetext{g}{Pleiades member (distance $\sim$ 133 pc).}
\tablenotetext{h}{See appendix for additional comments.}
\end{deluxetable}
\clearpage

\clearpage
\begin{deluxetable}{lcccl}

\tablecaption{Gliese ``Stars'' not in Table \ref{mainTable}\label{nonstars}}
\tablehead{
\colhead{Name }		&	\colhead{Alt name}	&	\colhead{RA}	         &	\colhead{Dec}            & \colhead{Reason not in Table \ref{mainTable}}	\\
  &  &	\colhead{J2000}	&	\colhead{J2000}	& 	}
\startdata	
Gl 127.1 B &	           & &		& Companion does not exist$^a$\\
Gl 452  B  & LHS 2470a	   & &		& Only an optical companion$^a$\\
Gl 678.1 B &  Wo 9592 B    & 17 30 21.96 & +05 33 06.5 & Only an optical companion \\
Gl 732  B  &               & &		& Companion does not exist$^a$\\
Gl 863.1 C & Wo 9788 C     & &		& Only an optical companion \\
GJ 1154 B  &               & 12 14 18.17 & +00 37 29.7 & Only an optical companion \\
GJ 2087	   & LHS 6219      & &		& Probable plate flaw$^a$\\
GJ 3063	   &               & &		& Unable to recover this star$^a$\\  
GJ 3750	   & QSO B1246+586 & 12 48 18.78 & +58 20 28.7 & BL Lac object$^b$ \\
GJ 3848	   & 7C 1424+2401  & 14 27 00.39 & +23 48 00.0 & BL Lac object$^b$ \\
GJ 4103    &               & &          & Unable to recover this star$^a$\\  
\enddata	
\tablenotetext{a}{See appendix for details.}
\tablenotetext{b}{Fleming et al.\ 1993}		
\end{deluxetable}

\clearpage
\begin{deluxetable}{lrrrrrl}
\tablecaption{Outliers in the 2MASS Color-Color Distribution
(Figure \ref{2massColors})
\label{outliers}}
\tablehead{
\colhead{Name }	& \colhead{$J-H$} & \colhead{$H-K_s$}	
  & \colhead{$J$}   & \colhead{$H$}   & \colhead{$K_s$}
  & \colhead{Comment} }
\startdata	
Gl 166 B & -0.137 & 0.125 & 9.849 & 9.986 & 9.861 &   white dwarf; 8\arcsec\ away from C, which is 3 mag brighter \\
Gl 168.1 & 0.784 & 0.239 & 11.833 & 11.049 & 10.810 &  \\
Gl 425 B & 0.779 & 0.033 & 6.638 & 5.859 & 5.826 &   4\arcsec\ binary \\
Gl 552.1 & 1.171 & 0.153 & 8.597 & 7.426 & 7.273 &   2MASS $J$ is bad; see Leggett (1992) \\
Gl 618 B & 0.304 & 0.270 & 8.339 & 8.035 & 7.765 &   8\arcsec\ binary; B much fainter than A \\
Gl 752 B & 0.682 & 0.461 & 9.908 & 9.226 & 8.765 &   VB10; very late-type M dwarf\\
GJ 1230 B & 0.829 & 0.257 & 8.860 & 8.031 & 7.774 &   5\arcsec\ dM-dM binary  \\
GJ 2110 & 0.815 & 0.273 & 7.434 & 6.619 & 6.346 &   Distant giant ($>$100 pc) \\
GJ 2129 & 0.913 & 0.258 & 8.131 & 7.218 & 6.960 &   Distant giant ($>$100 pc) \\
GJ 3107 & 0.856 & 0.273 & 8.434 & 7.578 & 7.305 &   Probably a giant \\
GJ 3140 & 0.903 & 0.313 & 8.287 & 7.384 & 7.071 &   Distant giant ($>$100 pc) \\
GJ 3231 B & 0.639 & -0.035 & 8.029 & 7.390 & 7.425 &  2MASS $K_s$ flagged as non-optimal; 20\arcsec\ binary \\
GJ 3276 & 0.990 & 0.491 & 12.757 & 11.767 & 11.276 &   LH 190 (probably Hyad); faint late-type M dwarf\\
GJ 3364 & 0.347 & 0.393 & 11.659 & 11.312 & 10.919 &   \\
GJ 3361 B & 0.533 & -0.077 & 6.661 & 6.128 & 6.205 &   5\arcsec\ binary\\
GJ 3517 & 0.743 & 0.527 & 11.212 & 10.469 & 9.942 &   LHS 2065; very late-type M dwarf\\
GJ 3655 & 0.695 & 0.498 & 11.928 & 11.233 & 10.735 &   LHS 2397A; very late-type M dwarf\\
GJ 3849 & 0.765 & 0.481 & 11.990 & 11.225 & 10.744 &   LHS 2924; very late-type M dwarf\\
GJ 4370 & 0.141 & 0.512 & 8.841 & 8.700 & 8.188 &    2MASS $J$ \& $K_s$ flagged as non-optimal\\
Wo 9383 A & 1.076 & 0.219 & 9.148 & 8.072 & 7.853 &   3.4\arcsec\ binary \\
Wo 9383 B & 1.236 & 0.205 & 9.366 & 8.130 & 7.925 &   3.4\arcsec\ binary \\
Wo 9584 C & 0.356 & 0.235 & 9.020 & 8.664 & 8.429 &    2MASS $K_s$ flagged as non-optimal \\
\enddata	
\end{deluxetable}

\end{document}